\documentclass[aip,rsi,reprint,graphicx]{revtex4-1}
\usepackage{ifpdf}
\usepackage{amsmath}
\usepackage{float}
\usepackage{subfigure}
\ifpdf 
    \usepackage[pdftex]{graphicx}   
    \usepackage[pdftex,     
            plainpages=false,   
            breaklinks=true,    
            colorlinks=true,
            pdftitle=Coil Project
            pdfauthor=D. Sabulsky
           ]{hyperref} 
    \usepackage{thumbpdf}
\else 
    \usepackage{graphicx}      
    \usepackage{hyperref}      
\fi 
\begin{document}
\begin{abstract}
We present the design, construction and characterization of Bitter-type electromagnets 
which can generate high magnetic fields under continuous operation with efficient heat 
removal for cold atom experiments. The electromagnets are constructed from a 
stack of alternating layers consisting of copper arcs and insulating polyester spacers. 
Efficient cooling of the copper is achieved via parallel rectangular water cooling 
channels between copper layers with low resistance to flow; 
a high ratio of the water-cooled surface area to the volume of copper ensures a short 
length scale $(\sim 1~\text{mm})$ to extract dissipated heat. High copper fraction per 
layer ensures high magnetic field generated per unit energy dissipated. The ensemble is 
highly scalable and compressed to create a watertight seal without epoxy. 
From our measurements, a peak field of $770$ G is generated 
$14$ mm away from a single electromagnet with a current of $400$ A and a total power 
dissipation of $1.6$ kW. With cooling water flowing at $3.8$ $l$/min, the coil temperature 
only increases by $7$ $^{\circ}$C under continuous operation. 
\end{abstract}
\pacs{07.55.Db, 07.55.-w, 85.70.Sq}
\title{Efficient Continuous-Duty Bitter-Type Electromagnets for Cold Atom Experiments} 
\affiliation{The James Franck Institute, Enrico Fermi Institute and Department of Physics, \\ The University of Chicago, Chicago, Illinois 60637, USA} 
\author{Dylan O. Sabulsky} 
\author{Colin V. Parker}
\author{Nathen D. Gemelke}
\thanks{Currently at the Department of Physics, The Pennsylvania State \\ University, University Park, Pennsylvania 16802, USA}
\author{Cheng Chin}
\thanks{Author to whom correspondence should be addressed. Electronic mail: \href{mailto:cchin@uchicago.edu}{cchin@uchicago.edu}} 
\date{\today}
\maketitle
\newpage
\section{Introduction}
Sustaining large, reliable magnetic fields is a major requirement across a broad range of 
applications in cold atom experiments, including Zeeman slowers\cite{Phillips, lafyatis,ketterle}, 
magnetic traps\cite{bergman,cornell,gucci,bloch}, magnetic conveyor belts\cite{hansch} and Feshbach 
control\cite{grimmchin}. All of these applications typically require magnetic fields on 
the order of 1,000 G or higher. The need to reach high fields has led to substantial 
evolution and innovation in the design of electromagnets for both resistive magnets and those based on 
superconducting elements\cite{bitter1,bitter2,bitter3,bruce}. 
In contrast to superconducting magnets, resistive magnets allow fast field control for 
the experiment, but exhibit significant heat dissipation;
heat removal is of critical importance in the optimization of magnet design\cite{lips,bow}.  

Most electromagnets used in cold atom experiments are constructed out of copper wire or refrigeration  
tubing\cite{joch,thomas2,dhotre} wound into cylinders. 
These designs are easy to manufacture and can easily fit around a vacuum component. 
There are many drawbacks to these designs, however. Electrical power is dissipated into cooling water that 
flows in series with the current, leading to a large temperature gradient in the device, 
a high resistance to cooling water, and significant differences in temperature between the 
water supply and return. Furthermore, construction of conventional coils requires a 
significant amount of epoxy, which can degrade over time, and the coil can not be easily modified. 
Finally, simple wire-wound electromagnets produce weaker magnetic fields 
by comparison to more advanced designs.
\begin{figure}[htbp]
\centering
\includegraphics[width=.45\textwidth]{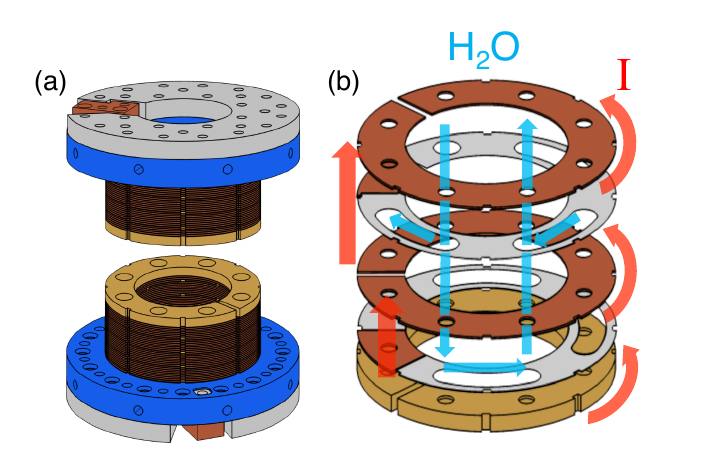}
\caption{Modified Bitter electromagnet design. (a) The steel mount (grey) and polycarbonate water distribution block (blue) are shown with mounting holes for fitting to a vacuum flange with a recessed viewport. Arc stacks consist of $31$ copper arcs (red) plus one brass arc (gold) as an end cap. (b) The current (red arrows) flows through the copper arcs (red) in a helical pattern through the device. The cooling water (blue arrows) flows parallel to the current through the device in channels cut into the polyester insulators (grey). For simplicity, we only show two copper arcs. The actual electromagnet contains $31$ copper arcs.}
\label{fig:1}
\end{figure}  

Recent investigations by Ricci \textit{et al} into electromagnet designs for cold 
atom experiments have focused on improved liquid cooling efficiency by introducing 
parallel cooling channels\cite{ricci}. The device supports current 
higher than that allowable in conventional designs; at $500$ A 
the device runs at $80$ $^{\circ}$C.
Another modified version of the wire-wound electromagnet design with a parallel cooling water 
channel within the coil has demonstrated significant improvement to the heat removal;
this device reaches a field of $1,500$ G while dissipating $6$ kW. 

Here we present a modified design for large bore Bitter type electromagnets to suit 
cold atom experiment, made watertight without epoxy. Each electromagnet delivers $770$ G 
while dissipating $1.5$ kW with modest machining and cooling requirements. The surface 
area of copper in contact with cooling water compared to the volume of copper is superior 
to copper wire and refrigeration tubing designs. A flow rate of $3.8$ $l$/min is obtained 
from a typical laboratory supply ($30$ psi or $207$ kPa), which keeps the heating of the 
electromagnet to a mere $7$ $^{\circ}$C, and a factor of up to $8$ times less heating than 
traditional and modern electromagnet designs. 
The design is suitable for high field tuning of Feshbach resonances\cite{grimmchin}, 
Zeeman slowing\cite{ketterle,lafyatis} and evaporation\cite{bloch,gucci} in 
cold atom experiments and related atomic physics applications. 

\section{The Modified Bitter Coil Design}
Our design follows the classic Bitter-type electromagnet \cite{bitter1}, consisting of $31$ 
annular copper arcs and one annular brass arc, separated by insulating polyester 
spacers of the same shape. In each successive layer, both copper and polyester 
layers are rotated by one-eighth of a turn to create a helical conduction path. 
The assembly is held together by seven steel screws and one brass screw bolted through a series of eight holes 
bored parallel to the coil axis. 
The brass screw serves to deliver electrical current to the annular brass arc 
and in turn to the first copper arc, furthest from the mount. The current returns through the
last copper arc. In order to ensure a clear optical path, the inner bore of the coil is significantly larger than more 
common Bitter coil designs of similar size\cite{bitter1}. The steel mount has threaded holes for optical mounts and mounting holes 
for attachment to our vacuum chamber. Our electromagnets reside inside recesses surrounding the viewports
of our vacuum chamber. Two independent electromagnets on opposite sides of the 
vacuum chamber allow for either a nearly homogenous field in a parallel current 
configuration or a quadrupole field in an antiparallel current configuration.

\begin{figure}[htbp]
\centering
\includegraphics[width=.5\textwidth]{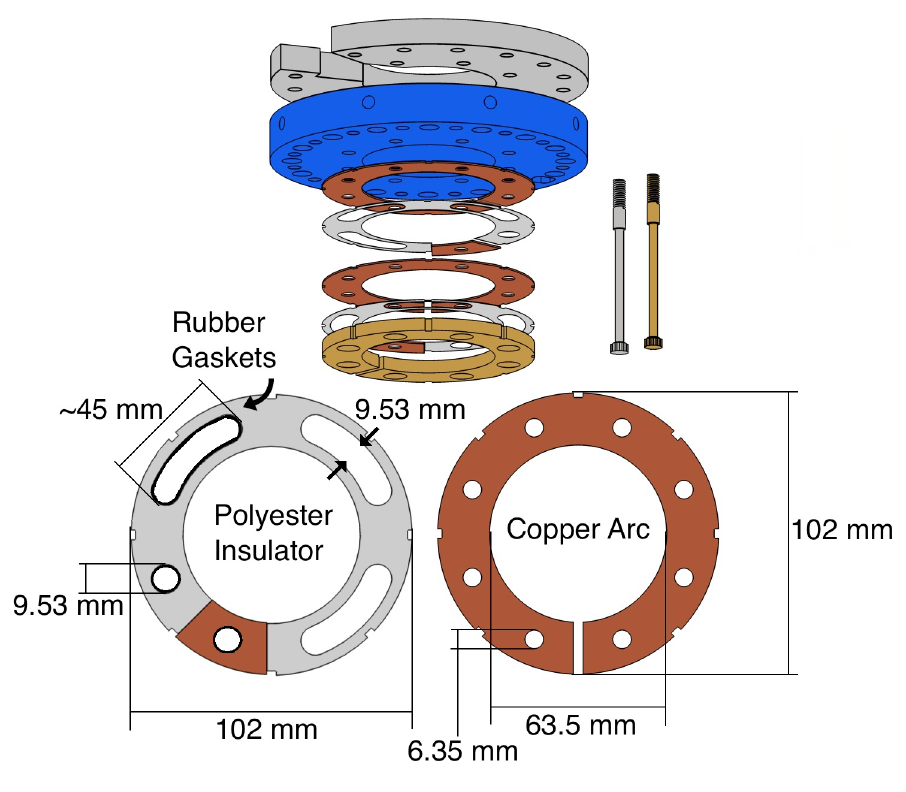}
\caption{Bitter electromagnet breakout diagram. Each arc has eight holes bored along the circumference to deliver cooling water and accommodate seven steel screws and one brass screw. Polyester spacers are similarly formed, with an additional wedge removed to allow electrical contact between copper arcs (red) through a copper insert (red) in the shape of the wedge. The polyester has channels cut to allow the passage of water. These channels are lined with nitrile gaskets (black). Fluid is delivered and extracted from alternating vertical channels in order to maximize fluid conductance and minimize the serial path taken by the cooling fluid. Only two copper arcs are shown for simplicity.}
\label{fig:2}
\end{figure}

We deviate from the conventional Bitter-type design by using eight bores to 
deliver and remove cooling water to a series of ducts with rectangular cross-sections 
across the copper surface. An advantage of this design geometry over the original Bitter coil 
is that all electrical and cooling water feed throughs leave the device on the same side 
of the electromagnet instead of entering one side and exiting the other\cite{bitter2}, see Fig.~\ref{fig:1}. 
These water channels between copper layers are defined by removed material from the polyester spacers. 
The channels in the polyester spacers are rimmed with nitrile gaskets to ensure a watertight seal. A complete turn consists of 
a copper arc, a polyester spacer, a copper insert and five nitrile  gaskets, see Fig.~\ref{fig:2}. 
Copper arcs are $1$ mm in thickness and the filling fraction of copper in a complete turn is $\lambda=77\%$. 
  
Coil elements were machined using standard milling machines and machine lathes. 
Copper arcs were formed from $1$ mm sheets of ultra-pure (99.99\%) ultra-conductive copper alloy 
$101$, cut into $102$ mm squares, stacked and then clamped between two steel 
discs. Arcs were machined to the described form as a single stack. 
The mount and cooling water distribution elements were constructed from 
steel alloy $316$ and polycarbonate, respectively. The steel mount provides a rigid seat for the steel clamping screws. 
Nitrile  gaskets and polyester insulators were cut with a CO$_{2}$ laser cutter 
(Universal VLS4.60). The entire device is sealed by physical compression to greater than $30$ psi above atmosphere without chemical sealants.
The steel screws have acetal 
washers to prevent shorts to the annular brass arc; 
this isolates the whole steel mount from the coil with a measured resistivity over 
$17$ k$\Omega$. A schematic of the electromagnet is shown in Fig.~\ref{fig:2} and a picture of the coil is shown in its test scaffolding in Fig.~\ref{fig:3_1}. 

It is useful to know how cooling water flows through the channels. The cross 
section of the rectangular ducts is $A=ab=1.6$ mm$^{2}$ where $a=0.254$ mm 
is the height and $b=6.35$ mm is the width. Each channel length $\ell$ is about $45$ mm.
Each insulator has three channels and the insulator between the last copper arc and 
the polycarbonate cooling block has four channels. This gives our design a total of 
$97$ channels with a total flow of $3.8$ $l$/min, making the average velocity of cooling 
water in a channel $U=0.4$ m/s. The channel height is less than the laminar boundary 
layer of fluid flow over a flat plate characterized by the displacement thickness\cite{munson} 
\begin{equation}\label{} \delta \approx 4.91\sqrt{\frac{\nu \ell}{U}}\approx 1.55  ~\text{mm,} \end{equation}
where $\nu=8.9 \times 10^{-7}$ m$^{2}$/s is the kinematic viscosity for water. The Reynolds number, $R_{e}$,  
for the case of a wide rectangular duct is 
\begin{equation}\label{} R_{e}= \frac{Q d_{h}}{\nu A}=223, \end{equation}
where $Q\approx 6.5\times 10^{-7}$ m$^{3}$/s is the volumetric flow rate of $F=3.8$ $l$/min of cooling water and 
$d_{h}=\frac{2ab}{a+b}=0.48$ mm is the hydraulic diameter for our rectangular ducts.
For a Reynolds number of $R_{e}\approx 400$ the flow would be turbulent\cite{peng}, 
implying that the flow along the length of the cooling 
channels is in the laminar flow regime. Laminar flow induces less vibrational 
noise. To match the cooling water flow of our Bitter electromagnets, refrigeration 
tubing and flat wire designs would need to run at very high water pressure 
on the order of $160$ psi or $1.1$ MPa, which is inconvenient and hazardous.
Further, our use of parallel cooling channels and the fact that flow 
conductance for this design increases roughly in proportion to the total number 
of turns results in a highly scalable design. With copper tubing wound 
into a cylindrical electromagnet, increasing the number of turns would cause a 
proportional increase in the hydrodynamic impedance. 

We investigate two ways to characterize the thermal characteristics of the 
electromagnet. Each cooling channel touches two copper arcs which leads to a large 
surface area of copper in contact with cooling water. The ratio of the surface area of 
copper in contact with cooling water to the volume of copper for the electromagnet is 
\begin{equation}\label{} \frac{\text{Cooling area}}{\text{Volume}}=1.0~\text{mm}^{-1}, \end{equation}
\noindent which suggests that dissipated heat in the coil can be efficiently removed 
within an average propagation distance of merely $1.0$ mm.    
We can further characterize the efficiency to remove heat by the Nusselt number, $N_{u}$. 
For laminar flow in a duct with a rectangular cross-section we obtain\cite{hoffman} 
\begin{equation}\label{} N_{u} \approx 1.94+0.659 \frac{b}{a}-0.02496 \bigg(\frac{b}{a}\bigg)^{2}=2.8. \end{equation}
The Nusselt number parameterizes the dominance of convective over conductive flow. 
A Nusselt number of $100$ to $1000$ corresponds to heat removal by convective flow, 
implying the heat removal of our device comes mostly from conductive flow.   

\section{Operational Profile}  
\begin{figure}
    \centering
    \hspace{-8cm}(a)
    
    \subfigure
    {\includegraphics[width=.315\textwidth]{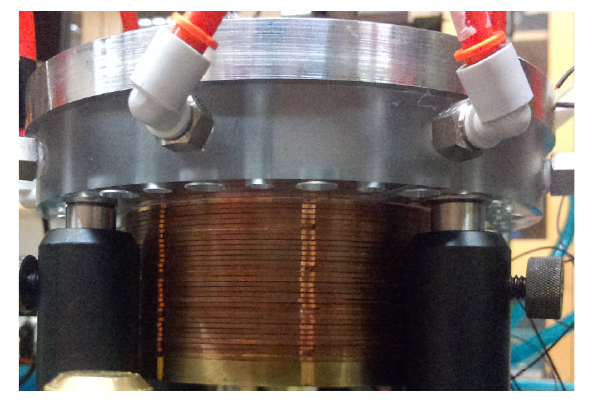}
     \label{fig:3_1}}
    \\
    \hspace{-8cm}(b)
    
    \subfigure
    {\includegraphics[width=.45\textwidth]{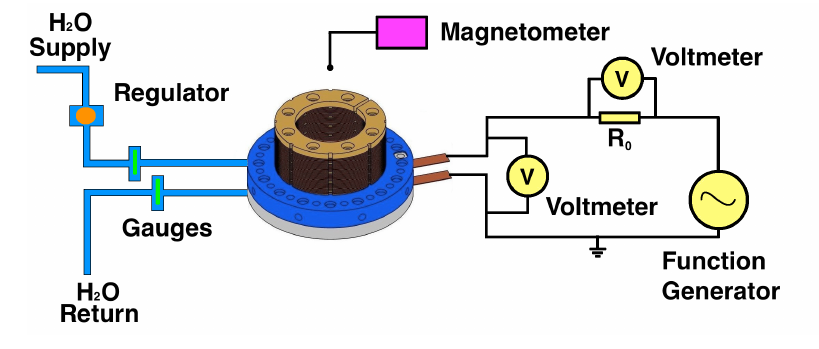}
     \label{fig:3_2}}
     \caption{Photo of a completed electromagnet in test scaffold, (a), and block diagram of water cooling, magnetic field and electrical measurement setup, (b). Hydraulics displayed on the left. Water flow is controlled and measured by a regulator (Watts H560) and a flowmeter (Blue-White Industries F-450 polysulfone tube rotameters) at the laboratory supply and monitored by pressure gauges (Watts DPG-3) on the supply and return immediately outside of the coil. Electrical tests shown schematically on the right. We feed the coil directly from a high current power supply (Agilent $6690$A). We measure the voltage drop across the coil and a resistor, $R_{0}=0.995$ $\Omega$, by use of a multimeter (Agilent 34401A); we do not use the resistor in high current testing. The temperatures of the coil, the brass arc, the steel mount and the copper connector are monitored with thermistors. The axial component of the magnetic field was measured by a magnetometer (Alphalabs GM-2) where the head was placed $14$ mm away from the base plate of the coils which approximates the distance to where atomic clouds will be in our experiment. An arbitrary function generator (SRS DS345) tests the AC response of the electromagnet.}
\end{figure}
In order to obtain an operational profile for the coils, we use the setup 
shown in Fig.~\ref{fig:3_2}. We measure the magnetic field, the resistance, 
the inductance and thermal characteristics. Magnetic field measurements and 
thermal characteristics were obtained together during electrical stress tests. 
The resistance and inductance were measured separately. 
\begin{figure}
    \centering
     \hspace{-8cm}(a)
    
    \subfigure
    {\includegraphics[width=.45\textwidth]{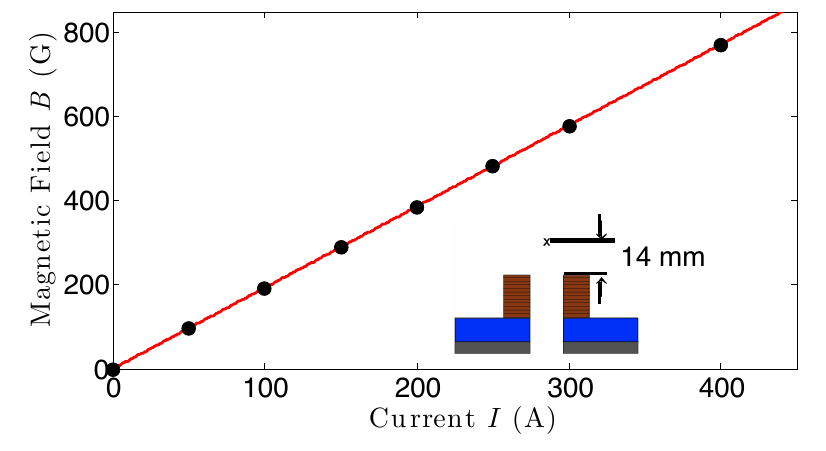}
     \label{fig:4_1}}
    \\
    \hspace{-8cm}(b)
    
    \subfigure
    {\includegraphics[width=.45\textwidth]{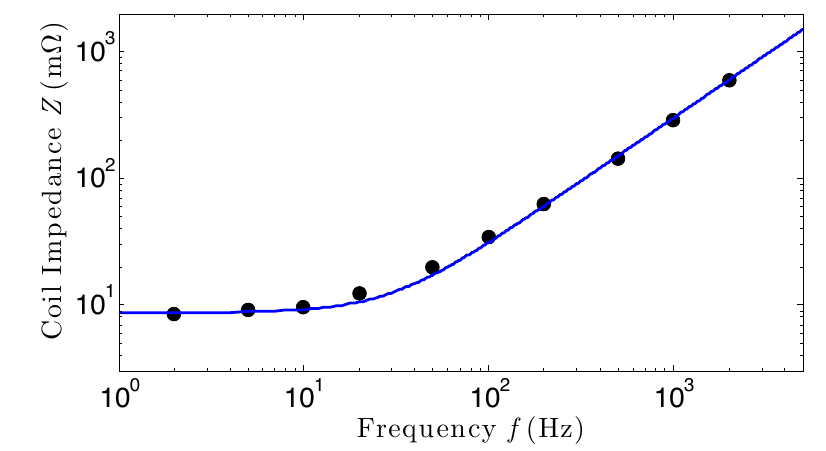}
     \label{fig:4_2}}
     \caption{Characterization of the Bitter electromagnet. (a) We measured the external magnetic field as a function of applied current $14$ mm above an electromagnet. The solid line is a linear fit yielding $1.93(2)$ G/A. (b) The inductance was calculated from measuring the impedance of the coil as a function of frequency and fit to the equation $Z(f)=(R^{2}+4\pi^{2}f^{2}L^{2})^{1/2}$, where $f$ is the frequency, $L=47.4(1)$ $\mu$H and resistance $R=8.7(1) ~\text{m}\Omega$ is determined with a DC current $I=10$ A.}
\end{figure}

To experimentally 
investigate the external magnetic field, we increased the current from $I=0$ A to $400$ A 
in $50$ A intervals over $20$ minutes. This allowed us to monitor the temperature in 
order to determine stability of the operating mode. The axial magnetic field $14$ mm away 
from the end cap of the coil was found to be $1.93(2)$ G/A, see Fig.~\ref{fig:4_1}. 
Application of $400$ amps would produce a peak field of $1,540(20)$ G for a 
pair of coils separated by $28$ mm. To predict the field, we start with the averaged current 
density $j(r)$ of a Bitter electromagnet, given by 
\begin{equation}\label{} j(r)=\frac{I}{d \lambda \ln{(a_{2}/a_{1})}}\frac{1}{r},\end{equation}
where $r$ is the radial position, $a_{1}=31.75$ mm is the inner radius of the electromagnet, 
$a_{2}=50.80$ mm is the outer radius, $\lambda=0.77$ is the copper filling factor and $I$ is the applied current. 
The $1/r$ current distribution results from the cylindrical geometry of the 
electromagnets. The magnetic field generated by a single copper arc along the 
axial direction can be expressed as\cite{qui} 
\begin{equation}\label{} B_{z}(z)=\frac{\mu_{0}d\lambda}{2}\bigg[ j\bigg(\sqrt{a^{2}_{1}+z^{2}}\bigg)-j\bigg(\sqrt{a^{2}_{2}+z^{2}}\bigg) \bigg], \end{equation} 
where $\mu_{0}=4\pi \times 10^{-7}$ H/m is the magnetic permeability and $z$ is the position from the center 
of the arc. This equation predicts that we should output $1.87$ G/A for our electromagnet 
which agrees well with our measurement of $1.93(2)$ G/A. 

The electrical response of the Bitter electromagnet is measured by the set up 
shown in Fig.~\ref{fig:3_2}. The resistance can be estimated by 
\begin{equation}\label{} R=\frac{2 \pi N \rho_{0}}{d\lambda \ln{(a_{2}/a_{1})}}, \end{equation}
where $N=31$ is the number of turns, $d=1$ mm is the thickness of a copper arc and $\rho_{0}=1.7\times 10^{-8}$ $\Omega$m is the resistivity of copper. 
This equation predicts a resistance of $R=9.0$ m$\Omega$. This is in agreement with our measured value of 
$R=8.7(1)$ m$\Omega$. 

We measure the impedance of the coil over a 
frequency range of $f=2$ Hz to $2$ kHz subjected to AC voltage at $4$ V peak to peak 
and extract the inductance $L$, see Fig.~\ref{fig:4_2}. For 
the inductance we obtained $L=47.4(1)$ $\mu$H. 
The predicted Bitter coil inductance is expressed by 
\begin{equation}\label{}  L=\frac{\mu_{0}N^{2}}{4\pi}a_{1}\Lambda\bigg(\frac{a_{2}}{a_{1}},\frac{t}{a_{1}}\bigg), \end{equation}
where $t$ is the half height of the electromagnet and $\Lambda$
is the self-inductance factor which depends on the geometry. The 
dimensionless factor of self-inductance $\Lambda$ depends only on the current 
distribution and geometry of the electromagnet. This factor for our Bitter 
electromagnets was found to be $\Lambda\approx 15$, 
compared to the typical Bitter electromagnet with $\Lambda\approx 30$ from literature\cite{krat}. 
The calculated inductance of $L=45.8$ $\mu$H is in good agreement with our measurement 
of $L=47.4(1)$ $\mu$H. Our device exhibits an inductance at least two orders of magnitude 
below conventional designs. The Bitter electromagnets can thus respond faster.    
\begin{figure}[htbp]
\centering
\includegraphics[width=.49\textwidth]{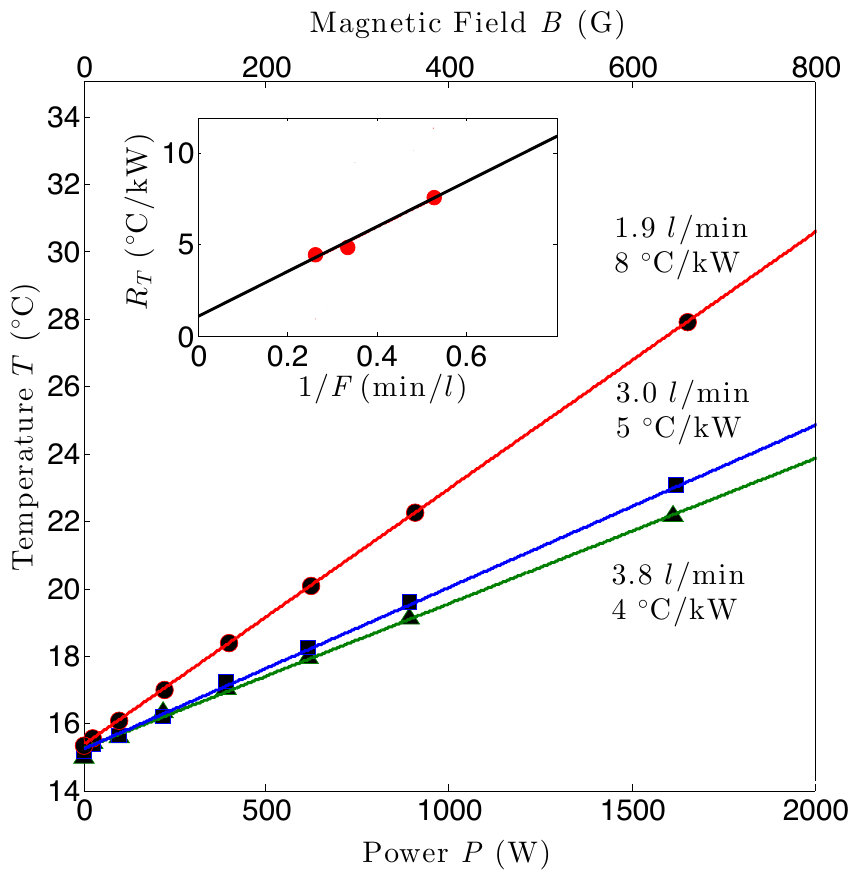}
\caption{Coil temperature as a function of current dissipated across the coils. We record the temperature of the copper arcs $T$ as we increase the current through the electromagnet. We perform this measurement for various flow rates to determine the dependance of coil temperature on the flow rate. Inset: dependence of thermal resistance $R_{T}=\Delta T/P$ on the inverse water flow rate $1/F$.}
\label{fig:5}
\end{figure} 

To measure how the power dissipation varies with the 
water flow we took multiple electrical tests at three different flow rates.  
We observe a linear dependance of the coil temperature on the dissipated power. 
The change in temperature of the coil, $\Delta T$, as a function of the 
flow rate $F$ and dissipating power $P$ is\cite{leanhard,zickler} 
\begin{equation}\label{} \Delta T=\frac{P}{\rho C F}=R_{T}P, \end{equation}
where $R_{T}=\Delta T/P$ is the absolute thermal resistance, the 
water heat capacity is $C=4186$ J/kg C$^{\circ}$ and density is $\rho=1.0$ g/cm$^{3}$. 
The thermal resistance $R_{T}$ of the tests ranges from $4$ $^{\circ}$C/kW 
to $8$ $^{\circ}$C/kW and depends inversely upon the flow rate $F$, see Fig.~\ref{fig:5}. 
We found the slope of the thermal resistance versus the inverse flow rate to be 
$0.20$ $^{\circ}$C cm$^{3}$/W s which is in good agreement with the predicted value, 
$1/\rho C=0.23$ $^{\circ}$C cm$^{3}$/W s. The thermal resistance of the device at a given 
water pressure is of particular merit; during continuous use, for every kilowatt, the device only 
gains a few degrees Celsius. Conventional devices dissipate more power
at comparable amperage and further exacerbate this difference through serial 
cooling schemes. Both the thermal resistance and water resistance of our Bitter 
electromagnet are very low.

The predicted failure mode for continuous operation of a 
pair of symmetric coils is one 
in which the cooling water boils within the devices, which may force cracks to open in the seals. 
Based on the above measurements, this mode is predicted to occur around a power dissipation of 
$P \approx 22.4$ kW  and a current of $I=1.6$ kA per coil with cooling water flowing at 
$F=3.8$ $l$/min. The peak magnetic field along the axis of symmetry of the two coils, when 
separated by $28.0$ mm, is $B_{z}=6,200$ G.

In summary, we describe a simple, scalable resistive Bitter electromagnet design suitable 
for high-field cold atom experiments. Our design provides stable temperatures close to the 
ambient during high field operation. Our electromagnet significantly outperforms conventional 
electromagnets made of refrigeration tubing or flat wire due to high copper filling 
fraction and efficient heat removal. Further, our modified 
bitter electromagnet exhibits an inductance at least two orders of magnitude 
below that of other electromagnet designs. Modified elements from the original Bitter electromagnet design 
include tapped holes for optics, all feedthroughs on one side of the device, and a large central bore for clear 
optical access, which are essential for cold atom experiments. This design requires basic machining skills and readily available 
materials to produce and scale up. 

\vspace{.5cm}
\section*{Acknowledgments}

The authors wish to thank L. Clark, E. Hazlett, K. Jim\'{e}nez-Garc\'{\i}a, J. Johansen, P. Scherpelz and S.K. Tung for insightful conversation 
and H. Krebs for machining advice. The authors acknowledge support from the NSF-MRSEC program, NSF
Award No. PHY-1206095, AFOSR-MURI, and ARO-MURI 63834-PH-MUR.

  

\end{document}